\documentclass[aps,pra,twocolumn,showpacs,superscriptaddress]{revtex4-1}

\usepackage{graphicx}
\usepackage{amsmath}
\usepackage{amssymb}

\newcommand{\be}{\begin{equation}}
\newcommand{\ee}{\end{equation}}
\newcommand{\ba}{\begin{eqnarray}}
\newcommand{\ea}{\end{eqnarray}}

\newcommand{\br}{{\mathbf{r}}}

\newcommand{\vect}[1]{\mathbf{#1}}

\newcommand{\nn}{\nonumber}

\newcommand{\bas}{\begin{eqnarray*}}
\newcommand{\eas}{\end{eqnarray*}}
\newcommand{\iaparam}{\tilde{g}_{\mathrm{tot}}}

\begin{document}

\setlength{\arraycolsep}{1.5pt}

\title{Exotic vortex lattices in two-species Bose--Einstein condensates}
\date{\today}
\author{Pekko Kuopanportti}\email{pekko.kuopanportti@aalto.fi}
\author{Jukka A. M. Huhtam\"aki}
\affiliation{Department of Applied Physics/COMP, Aalto University, P.O. Box 13500, FI-00076 AALTO, Finland}
\author{Mikko M\"ott\"onen}
\affiliation{Department of Applied Physics/COMP, Aalto University, P.O. Box 13500, FI-00076 AALTO, Finland}
\affiliation{Low Temperature Laboratory, Aalto University, P.O. Box 13500, FI-00076 AALTO, Finland}

\begin{abstract}
We numerically investigate vortex lattices in rotating two-component Bose--Einstein condensates in which the two components have unequal atomic masses and interact attractively with each other. For sufficiently strong attraction, the system is found to exhibit exotic ground-state structures in a harmonic trap, such as lattices having a square geometry or consisting of two-quantum vortices. The obtained states satisfy the Feynman relation, and they can be realized with current experimental techniques.
\end{abstract}

\maketitle

\section{Introduction}

One of the characteristic properties of superfluids is that they respond to rotation by forming quantized vortices~\cite{Ons1949.NCim6Sup2.249,Fey1955.PLTP1.17}. Perhaps the most convenient environment to controllably study the physics of vortices is provided by dilute Bose--Einstein condensates (BECs) of alkali-metal atoms~\cite{Mat1999.PRL83.2498,Mad2000.PRL84.806,Abo2001.Sci292.5516}. The experiments with dilute BECs have verified Abrikosov's original prediction~\cite{Abr1957.ZETF32.1442} that when a condensate without spin degrees of freedom is set into rapid rotation, a triangular lattice of single-quantum vortices is created. The areal density of vortices $n_\mathrm{v}$ is given by the so-called Feynman relation~\cite{Fey1955.PLTP1.17}
\be\label{eq:Feynman}
n_\mathrm{v} = \frac{m\Omega}{\pi\hbar},
\ee
where $m$ is the mass of the constituent boson and $\Omega$ is the angular frequency at which the superfluid is being rotated. Equation~(\ref{eq:Feynman}) is obtained by assuming that, on average, the superfluid rotates like a rigid body---even though the superfluid flow is inherently irrotational and has a nonvanishing curl only at the singular vortex cores.

Vortex physics becomes much more diverse in BECs that consist of more than one component~\cite{Kas2005.IJMPB19.1835}. Even for the simplest multicomponent case, the two-component BEC, a rich variety of exotic ground-state vortex structures have been found in theoretical studies: coreless vortices that can be interpreted as skyrmions or meron pairs using a pseudospin representation \cite{Kas2004.PRL93.250406,Kas2005.PRA71.043611}, serpentine vortex sheets~\cite{Kas2009.PRA79.023606}, giant skyrmions~\cite{Yan2008.PRA77.033621,Mas2011.PRA84.033611}, and interlacing square vortex lattices~\cite{Mas2011.PRA84.033611,Mue2002.PRL88.180403,Kas2003.PRL91.150406,Kec2006.PRA73.023611,Min2009.PRA79.013605,Huh2009.PRA80.051601,Dah2008.PRB78.144510}. Of these, the singly quantized coreless vortices~\cite{Mat1999.PRL83.2498} and the square lattices~\cite{Sch2004.PRL93.210403} have already been observed experimentally. To date, two-component BECs have been realized in various setups: a single isotope in two different hyperfine spin states~\cite{Sch2004.PRL93.210403,Mya1997.PRL78.586,Hal1998.PRL81.1539,Mat1999.PRL83.2498,Del2001.PRA63.051602,And2009.PRA80.023603}, two different isotopes of the same alkali metal~\cite{Bap2008.PRL101.040402,Sug2011.PRA84.011610}, or two distinct elements~\cite{Fer2002.PRL89.053202,Mod2002.PRL89.190404,Cat2008.PRA77.011603,Tha2008.PRL100.210402,Aik2009.NJP11.055035,Ler2011.EPJD65.3}.

All the aforementioned vortex structures pertain to systems where the two components interact repulsively with each other. Indeed, the tendency to form structures where one component fills the vortex core of the other component is understood in terms of the minimization of the repulsive density--density interaction energy. Also the transition from triangular to square vortex lattices is driven by the repulsive intercomponent interaction~\cite{Mas2011.PRA84.033611,Mue2002.PRL88.180403,Kas2003.PRL91.150406,Kec2006.PRA73.023611}.

However, vortex lattices in rotating two-component BECs with \emph{attractive} intercomponent interaction have not been studied as thoroughly~\cite{Bar2008.NJP10.043030,Bar2010.NJP12.043004}. When all particles in the system have equal mass and experience the same external rotation, the two components are expected to form identical vortex lattices and behave in most respects like a single-component BEC. If the atomic masses of the two components are unequal, the situation becomes more interesting: On one hand, Eq.~(\ref{eq:Feynman}) implies that the vortex densities in the two components should differ from one another. The intercomponent attraction, on the other hand, results in an effective attraction between vortices in different components, favoring configurations in which the two vortex lattices have maximal overlap. Clearly, differently spaced triangular lattices can overlap without frustration only for a very limited range of atomic mass ratios. Thus, we pose the following questions: What is the combined effect of the mass difference and the intercomponent attraction on the arrangement of vortices in the rotating two-component system? Is the Feynman relation still obeyed? 

In this article, we aim to answer the above questions. We demonstrate numerically that the two-component BEC with unequal atomic masses and attractive intercomponent interaction supports exotic ground-state vortex configurations in a rotating harmonic trap. Depending on both the ratio of atomic masses and the strength of the intercomponent attraction, the ground states may contain overlapping square lattices or even lattices consisting of doubly quantized vortices. We find no significant deviations from the Feynman relation. Since two-atomic-species BECs have already been experimentally realized~\cite{Fer2002.PRL89.053202,Mod2002.PRL89.190404,Cat2008.PRA77.011603,Tha2008.PRL100.210402,Aik2009.NJP11.055035,Ler2011.EPJD65.3}, 
also with tunable intercomponent interactions~\cite{Tha2008.PRL100.210402}, the exotic lattices should be observable with the current state of the art. The two-component system with unequal masses was first studied by Barnett \emph{et al.}~\cite{Bar2008.NJP10.043030,Bar2010.NJP12.043004}, but they considered only mass ratios near unity and thus did not detect any nontriangular vortex lattices.

The remainder of this article is organized as follows. In Sec.~\ref{sc:theory}, we present the zero-temperature mean-field theory of the two-component BEC and derive the Thomas--Fermi approximation that provides useful information about the ground-state structure. We also generalize the Feynman relation and discuss the expected lattice geometries for selected values of the atomic mass ratio. Section~\ref{sc:results} presents the numerically obtained ground-state vortex lattices. In Sec.~\ref{sc:discussion}, we summarize and discuss the main results of the work.

\section{Theory}\label{sc:theory}

We restrict our studies to the zero-temperature mean-field regime, where the two-component BEC is described in terms of two complex-valued order-parameter fields $\Psi_1$ and $\Psi_2$. The system is set into rotation about the $z$ axis with angular frequency $\Omega$ which is assumed to be the same for both components. For simplicity, we only consider situations in which the $z$ dependence of $\Psi_1$ and $\Psi_2$ can be factored out and all the relevant dynamics occur in the $x$ and $y$ directions. This reduction is accurate, e.g., in the regime of strong axial confinement, where the harmonic trapping frequencies in the $z$ direction are much greater than the trapping frequencies $\omega_1$ and $\omega_2$ in the radial direction. The two-dimensional Gross--Pitaevskii (GP) energy functional in the rotating reference frame is given by $E=\int \varepsilon(\br)d^2 r$, where the energy density is
\ba\label{eq:GPEF}
\varepsilon(\br) &=& \sum_{j=1}^2 \bigg( \frac{\hbar^2}{2m_j}|\nabla\Psi_j|^2 +\frac{1}{2}m_j\omega_j^2 r^2 n_j + \frac{1}{2}g_{j} n_j^2 \nonumber \\&& -\Omega \Psi_j^\ast L_z \Psi_j \bigg) + g_{12}n_1 n_2.
\ea
Here, $m_1$ and $m_2$ are the masses of the bosons in the two components, $r^2=x^2+y^2$, and $L_z=-i\hbar\left[\vect{e}_z \cdot\left(\br\times \nabla \right)\right]$ is the $z$ component of the angular momentum operator in the position representation. The order parameters are normalized such that $\int n_j(\br)d^2 r=N_j$, where $n_j=|\Psi_j|^2$ is the areal particle density and $N_j$ the total particle number of component $j\in\left\{1,2\right\}$. In Eq.~(\ref{eq:GPEF}), $g_{1}$ and $g_{2}$, which measure the intracomponent interaction strengths, are assumed to be positive throughout the paper, and $g_{12}$ is the intercomponent coupling constant for which we consider only negative values. The parameter $g_j$ ($g_{12}$) is proportional to the corresponding $s$-wave scattering length $a_j$ ($a_{12}$), with the exact dependence being determined by the trap along the $z$ direction~\cite{Note_trap}.

Minimizing the GP energy functional with respect to variations of $\Psi_1$ and $\Psi_2$, and introducing the chemical potentials $\mu_1$ and $\mu_2$ to fix the particle numbers $N_1$ and $N_2$, yields the coupled GP equations
\ba
&\Big[&-\frac{\hbar^2\nabla^2}{2m_j}+\frac{1}{2}m_j\omega_j^2 r^2+ g_j n_j \nonumber \\ &&   +g_{12}n_{3-j} -\Omega L_z -\mu_j \Big] \Psi_j(\br) = 0,\label{eq:GPE}
\ea
where $j\in\left\{1,2\right\}$. Equations~(\ref{eq:GPE}) form the backbone of our numerical analysis in Sec.~\ref{sc:results}.

\subsection{Thomas--Fermi approximation}

Let us consider the Thomas--Fermi (TF) limit corresponding to large particle numbers~\cite{Fet2009.RMP81.647,Cor2009.PRA80.033609}, where the spatial variation of the condensate density is negligible compared to the other terms in the energy functional. Thus, we approximate $-i\hbar\nabla \Psi_j = \hbar(\nabla S_j)\Psi_j - i\hbar \exp\left(iS_j \right)\nabla|\Psi_j| \approx m_j \vect{v}_j \Psi_j$, where $S_j=\mathrm{arg}\left(\Psi_j\right)$ and $\vect{v}_j=\hbar\nabla S_j / m_j$ denotes the superfluid velocity of component $j$. Consequently, Eq.~(\ref{eq:GPEF}) simplifies to
\ba
\varepsilon_\mathrm{TF}(\br) &=& \frac{1}{2}\sum_{j=1}^2  \Big[ m_j\left(\vect{v}_j-\vect{v}_\mathrm{rb} \right)^2 n_j  +m_j \tilde{\omega}^2_j r^ 2 n_j \nonumber \\ & & + g_{j} n_j^2 \Big] + g_{12}n_1 n_2,
\ea
where $\tilde{\omega}_j := \sqrt{ \omega_j^2 - \Omega^ 2}$ and $\vect{v}_\mathrm{rb}=\Omega\vect{e}_z \times \vect{r}$ corresponds to classical rigid-body rotation. For large BECs with many vortices, the superfluid flow closely mimics the rigid-body behavior, that is, $\vect{v}_j \approx \vect{v}_\mathrm{rb}$ outside the vortex cores, and the first term vanishes to leading order. Hence, we end up with
\be
\varepsilon'(\br) = \frac{1}{2}\sum_{j=1}^2 \left[m_j \tilde{\omega}_{j}^ 2 r^2 n_j + g_{j} n_j^2 \right] + g_{12}n_1 n_2,
\ee
which has exactly the same form as the TF energy functional for a nonrotated two-component system, but the confining potentials are altered to $m_j \left(\omega_j^2-\Omega^2\right)r^2/2$.

Variation of $\varepsilon'$ with respect to $n_1$ and $n_2$ with fixed particle numbers $\int  n_j d^2 r = N_j$ gives two coupled equations that can be solved for $n_1$ and $n_2$. When $g_{1}g_{2}-g_{12}^2 > 0$, the components are miscible, and we find
\be\label{eq:TFA}
n_j(r)=n_{j,0} \left(1-r^2/R^2_j\right),
\ee
where the peak densities $\left\{n_{j,0}\right\}$ and TF radii $\left\{R_{j}\right\}$ are given by
\ba
n_{j,0}^2 &=&\frac{N_j}{\pi}\frac{g_{3-j}m_j\tilde{\omega}_j^2-g_{12}m_{3-j}\tilde{\omega}_{3-j}^2}{g_1 g_2-g_{12}^2},\label{eq:n0} \\
R_j^4 &=& \frac{4 N_j}{\pi} \frac{g_1g_2-g_{12}^2}{g_{3-j}m_j\tilde{\omega}_j^2-g_{12}m_{3-j}\tilde{\omega}_{3-j}^2}.\label{eq:R_alpha}
\ea
The healing lengths, which can be used to estimate the vortex core radii, are defined as $\xi_j = \hbar/\left(2 m_j \mu_j'\right)^{1/2}$, where $\mu_{j}'=g_j n_{j,0}+g_{12}n_{3-j,0}$ denotes the chemical potential in the TF approximation.

In order to simplify the numerical analysis, it is convenient reduce the number of free parameters in the problem. Thus, we will fix the parameters pertaining to component 2, i.e., $\omega_{r,2}$, $g_2$, and $N_2$, in a way that is most suitable for studying how the vortices reorganize when the interaction strength $g_{12}$ is varied. To this end, we require the TF profiles to be identical,  $n_{1,0}=n_{2,0}=:n_{0}$ and $R_1=R_2=:R$, as well as equating the healing lengths, $\xi_1=\xi_2=:\xi$. According to Eqs.~(\ref{eq:n0}) and (\ref{eq:R_alpha}), these conditions are met when
\ba
\omega_{2}^2 & = & \Omega^2 + \rho^2 \left(\omega_{1}^2-\Omega^2\right), \nn \\
g_2 &=& \rho g_1 + \left( \rho - 1 \right) g_{12}, \label{eq:TF-scalings} \\
N_2 &=& N_1, \nn
\ea
where $\rho=m_1/m_2$ denotes the atomic mass ratio. Consequently, the TF expressions simplify to
\ba
 n_0^2 &=& \frac{m_1 N_1}{\pi} \frac{\omega_ 1^2- \Omega^2}{g_1+g_{12}}, \nn \\
R^4 &=& \frac{4 N_1}{\pi m_1} \frac{g_1+g_{12}}{\omega_1^2-\Omega^2}, \label{eq:TF-quantities} \\
\xi^4 & = & \frac{\pi\hbar^4}{4  m_1^3 N_1} \frac{1}{\left(g_1+g_{12}\right)\left(\omega_1^2-\Omega^2\right)}.\nn 
\ea
We note that the interaction strengths enter Eqs.~(\ref{eq:TF-quantities}) only through the combination $g_1+g_{12}$ which may be interpreted as measuring the total density--density coupling strength. Therefore, it will be convenient to parametrize the system in terms of $g_1+g_{12}$ and the ratio $\sigma := -g_{12}/g_1$, where $0\leq \sigma < 1$. The case $\sigma \geq 1$ corresponds to the collapse of the BEC under attractive forces.

\subsection{Feynman relation and vortex lattices}

For two-component BECs, the Feynman relation (\ref{eq:Feynman}) naturally generalizes to
\be\label{eq:Feynman2}
n_{\mathrm{v},j} = \frac{m_j\Omega}{\pi\hbar}.
\ee
It states that for a two-component system with unequal masses, the vortex densities are also unequal, satisfying $n_{\mathrm{v},1}=\rho n_{\mathrm{v},2}$. Without loss of generality, we will assume that $\rho\geq 1$. It should be noted that Eq.~(\ref{eq:Feynman2}) differs significantly from the treatment of Barnett \emph{et al.}~\cite{Bar2008.NJP10.043030,Bar2010.NJP12.043004} who equate the two vortex densities and allow for component-dependent rotation frequencies.

How does the difference in the vortex densities affect the geometry of the underlying lattice? The attractive intercomponent interaction favors states in which vortices in different components lie on top of each other, since it acts to maximize the overlap $\int n_1 n_2 d^ 2 r$ between the two components. However, two regular triangular lattices of single-quantum vortices can perfectly overlap only when the mass ratio can be expressed as $\rho=k^2+l^ 2+kl$, where $k,l \in\mathbb{N}$. Therefore, the two-component BEC may support unconventional lattice geometries when $\rho$ is far from such a value and the intercomponent attraction is sufficiently strong.

\begin{figure}[h!]
\includegraphics[
  width=74mm,
  keepaspectratio]{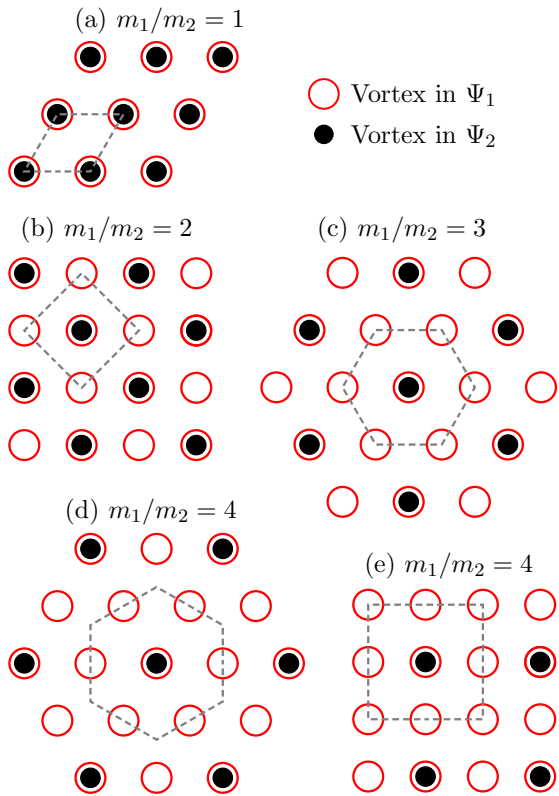}
\caption{\label{fig:lattices} (Color online) Vortex lattice geometries of the attractively interacting two-component BEC for different values of the mass ratio $\rho=m_1/m_2$; (a)~$\rho=1$; (b)~$\rho=2$; (c)~$\rho=3$; (d)--(e)~$\rho=4$. Hollow and solid circles represent vortices of components 1 and 2, respectively. The unit cell of the combined lattice is shown with dashed lines.
}
\end{figure}

The overlapping vortex lattices are illustrated in Fig.~\ref{fig:lattices}. For simplicity, we consider here only single-quantum vortices~\cite{Note_mqv} and integer values of $\rho$. In real experiments, $\rho$ is rarely strictly integer-valued but often sufficiently close to an integer such that the lattices in Fig.~\ref{fig:lattices} provide good approximations to the actual ones. For $\rho=1$ and $g_{12}<0$, the ground state consists of two superposed triangular lattices [Fig.~1(a)]. In the experimentally relevant case $\rho = 2$ (for ${}^{41}$K--${}^{87}$Rb BEC~\cite{Fer2002.PRL89.053202,Mod2002.PRL89.190404,Cat2008.PRA77.011603,Tha2008.PRL100.210402,Aik2009.NJP11.055035}, 
$\rho\approx 2.1$), a natural way to obtain $n_{\mathrm{v},1}/n_{\mathrm{v},2}=2$ is to arrange the vortices into two square lattices tilted by 45 degrees relative to one another [Fig.~\ref{fig:lattices}(b)]. For $\rho=3$, which applies to ${}^{41}$K--${}^{133}$Cs ($\rho\approx 3.2$) and ${}^{7}$Li--${}^{23}$Na ($\rho\approx 3.3$), triangular lattices are expected [Fig.~\ref{fig:lattices}(c)]. This is also the case for $\rho=4$ (in ${}^{23}$Li--${}^{87}$Rb mixtures, $\rho\approx 3.8$), as shown in Fig.~\ref{fig:lattices}(d). It is also possible to superpose two square lattices with $n_{\mathrm{v},1}/n_{\mathrm{v},2}=4$ [Fig.~\ref{fig:lattices}(e)], but such a state is expected to have a higher energy than the triangular arrangement due to the higher intracomponent energies. In the next section, we present numerical solutions of the two-component GP equations which exhibit the lattice geometries depicted in Fig.~\ref{fig:lattices}.

\section{Numerical results}\label{sc:results}

In order to demonstrate that the intercomponent interaction can induce a change in the ground-state vortex geometry, we have numerically solved the coupled GP equations of a rotating two-component BEC with $m_1\neq m_2$. We consider the mass ratios $\rho=m_1/m_2\in\left\{2,3,4\right\}$. In the numerical simulations, Eqs.~(\ref{eq:GPE}) are discretized on a uniform grid and solved using a relaxation method. We measure length in units of the radial harmonic oscillator length $a_r =\sqrt{\hbar/m_1\omega_1}$ and energy in units of $\hbar\omega_1$ and normalize the dimensionless order parameters to unity. The interactions are parametrized by the dimensionless quantities $\iaparam := \left(g_1 + g_{12}\right)N_1/\hbar\omega_1 a_r^2$ and $\sigma:=-g_{12}/g_1$. The values we choose for the total interaction parameter $\iaparam$ are large enough such that the TF expressions, Eqs.~(\ref{eq:TFA}) and~(\ref{eq:TF-quantities}), accurately approximate the numerically obtained states, thereby justifying our use of Eqs.~(\ref{eq:TF-scalings}). To emphasize the role of the intercomponent attraction, we consider a range of values for $\sigma$ in the interval $0 \leq \sigma < 1$. In experiments, the ratio $g_{12}/g_1$ may be controlled with intercomponent Feshbach resonances~\cite{Chi2010.RMP82.1225}, which have been demonstrated for various alkali-metal mixtures~\cite{Tha2008.PRL100.210402,Sta2004.PRL93.143001,Ino2004.PRL93.183201,Pil2009.PRA79.042718}.

\begin{figure}[h!]
\includegraphics[
  width=72mm,
  keepaspectratio]{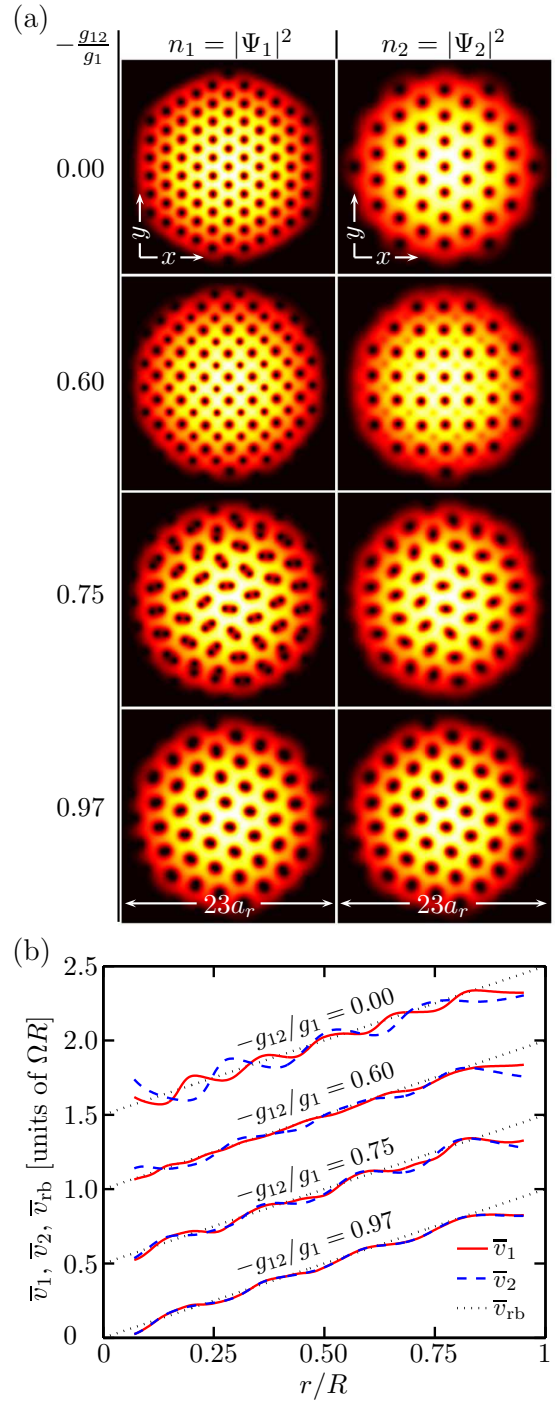}
\caption{\label{fig:rho2profiles} (Color online) Ground states of a two-component BEC with the mass ratio $m_1/m_2=2$, rotation frequency $\Omega=0.97\,\omega_1$, and interaction strength $\iaparam = 705$.  (a)~Particle densities $n_1$ and $n_2$ at different values of the intercomponent attraction strength $\sigma = -g_{12}/g_1$. (b)~Average azimuthal velocities $\overline{v}_{1}$ and $\overline{v}_{2}$ as functions of $r$ [Eq.~(\ref{eq:avevelo})] for the states shown in (a). For clarity, the curves have been shifted vertically; the dotted line $\overline{v}_{\mathrm{rb}}=\Omega r$ passes through the origin in all cases. The Thomas--Fermi radius $R=11.1\,a_r$, where $a_r=\sqrt{\hbar/m_1 \omega_1}$.}
\end{figure}

Figure~\ref{fig:rho2profiles}(a) presents the ground-state density profiles $n_1$ and $n_2$ for the mass ratio $\rho=2$, rotation frequency $\Omega = 0.97\,\omega_1$, total interaction parameter $\iaparam=705$, and different values of the intercomponent attraction strength $\sigma$. At $\sigma=0$, the components are noninteracting, and the ground state contains triangular lattices in both components. However, as the intercomponent attraction increases, the ground-state profiles change gradually into square lattices as shown for $\sigma=0.6$ [see also Fig.~\ref{fig:lattices}(b)]. A further increase in $\sigma$ eventually results in the triangular geometry presented in Fig.~\ref{fig:lattices}(a), but with vortex pairs occupying the lattice points in component 1 instead of solitary vortices. The vortex pairs in $\Psi_1$ overlap with single vortices in $\Psi_2$ that have elongated cores. Finally, at $\sigma=0.97$, the vortex pairs have merged to form a triangular lattice of two-quantum vortices in $\Psi_1$, while $\Psi_2$ contains an essentially identical arrangement of single-quantum vortices. Although we present results only for the values of $\Omega$ and $\iaparam$ mentioned above, qualitatively similar behavior is observed also for other values as long as the states contain a sufficient number of vortices. We point out that the square vortex lattice shown here for $\sigma=0.6$ also appears in a system of two homogeneous superfluids with optically induced current--current interactions and $\rho = 2$~\cite{Dah2008.PRL101.255301}.

To assess the validity of the Feynman relation, Eq.~(\ref{eq:Feynman2}), it is useful to investigate the velocity fields $\mathbf{v}_j = \hbar\nabla S_j/m_j$. We define the average azimuthal velocities as  
\be\label{eq:avevelo}
\overline{v}_j(r) = \frac{\int_0^{2\pi}   \mathbf{e}_\phi \cdot \mathbf{v}_j(r,\phi)n_j(r,\phi)d \phi}{\int_0^{2\pi}n_j(r,\phi) d \phi},
\ee
where $\phi$ is the polar angle. The average is weighted with the particle density to smooth out the velocity divergences at the singular vortex cores. For classical rigid-body rotation, the velocity field is $\mathbf{v}_\mathrm{rb}= \Omega \vect{e}_z \times \br$, and hence Eq.~(\ref{eq:avevelo}) gives $\overline{v}_\mathrm{rb}(r)=\Omega r$. If Eq.~(\ref{eq:Feynman2}) holds and the superfluid flow mimics the rigid-body behavior, $\overline{v}_1(r)$ and $\overline{v}_2(r)$ should approximately fall on top of the line $\Omega r$. In Fig.~\ref{fig:rho2profiles}(b), the velocities are plotted as functions of $r$ for the states shown in Fig.~\ref{fig:rho2profiles}(a). We observe that $\overline{v}_1$ and $\overline{v}_2$  closely follow the line $\Omega r$, in accordance with the Feynman relation, irrespective of the value of $\sigma$. When $\sigma$ is sufficiently large, $\overline{v}_1$ and $\overline{v}_2$ lie accurately on top of each other, reflecting the fact that the vortex cores in the two components overlap. 

\begin{figure}[ht!]
\includegraphics[
  width=72mm,
  keepaspectratio]{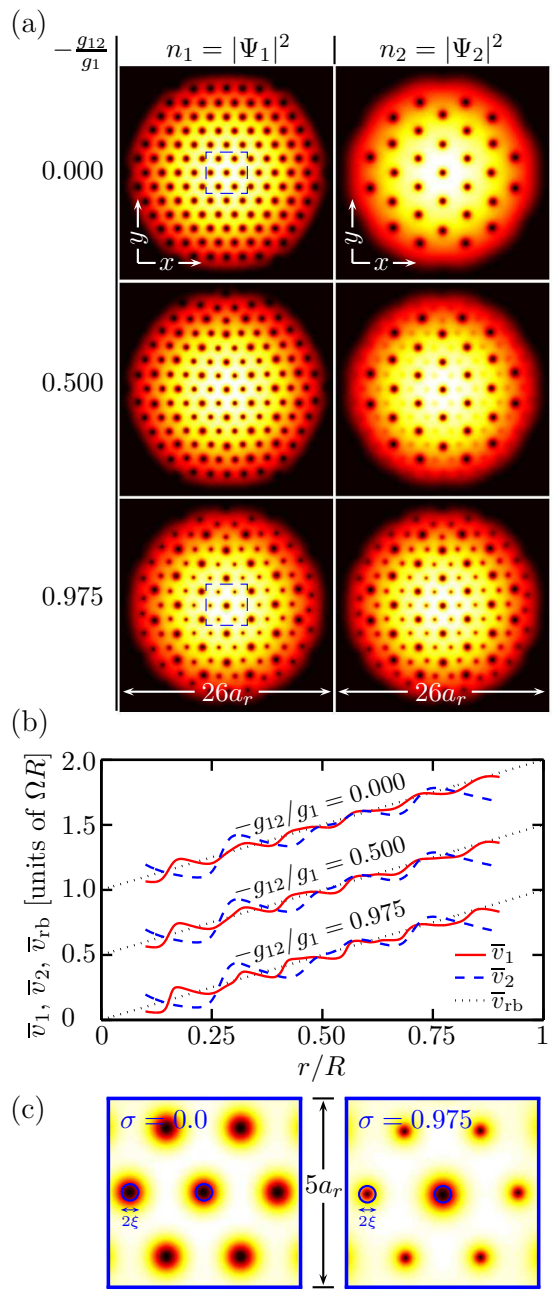}
\caption{\label{fig:rho3profiles} (Color online) Ground states of a two-component BEC with $m_1/m_2=3$, $\Omega=0.95\,\omega_1$, and $\iaparam = 2000$. (a)~Densities $n_1$ and $n_2$ at different values of $\sigma = -g_{12}/g_1$. (b)~Average azimuthal velocities $\overline{v}_{1}$ and $\overline{v}_{2}$ [Eq.~(\ref{eq:avevelo})] for the states presented in (a). The curves have been shifted in the vertical direction for clarity, and $R=12.7\,a_r$. (c)~Magnified view of $n_1$ for $-g_{12}/g_1=0$ and $0.975$ [boxed areas in (a)] illustrating the variation of the vortex core size. The circles have the radius $\xi=0.252\,a_r$. 
}
\end{figure}

For the mass ratio $\rho=3$, two triangular lattices can efficiently overlap [Fig.~\ref{fig:lattices}(c)], and consequently we have not observed any nontriangular geometries for large lattices, despite varying the parameters $\Omega$, $\iaparam$, and $\sigma$ extensively. Representative ground states for $\rho=3$ are depicted in Fig.~\ref{fig:rho3profiles}(a). The vortex lattices retain their triangular shape with increasing $\sigma$. However, those vortices in $\Psi_1$ that are paired with vortices in $\Psi_2$ increase in core size as $\sigma$ increases, whereas unpaired vortices become smaller. A comparison of the various vortex core sizes with the healing length $\xi$ is presented in Fig.~\ref{fig:rho3profiles}(c), where the core size at $\sigma=0$ is observed to lie between the core sizes of the paired and unpaired vortices at $\sigma=0.975$. Moreover, the unpaired vortices of component 1 are found to induce density craters in component 2 at sufficiently large values of the intercomponent attraction as shown in Fig.~\ref{fig:rho3profiles}(a) for $\sigma=0.975$. The azimuthal velocities plotted in Fig.~\ref{fig:rho3profiles}(b) again show no significant deviation from the Feynman relation.

\begin{figure}[ht!]
\includegraphics[
  width=72mm,
  keepaspectratio]{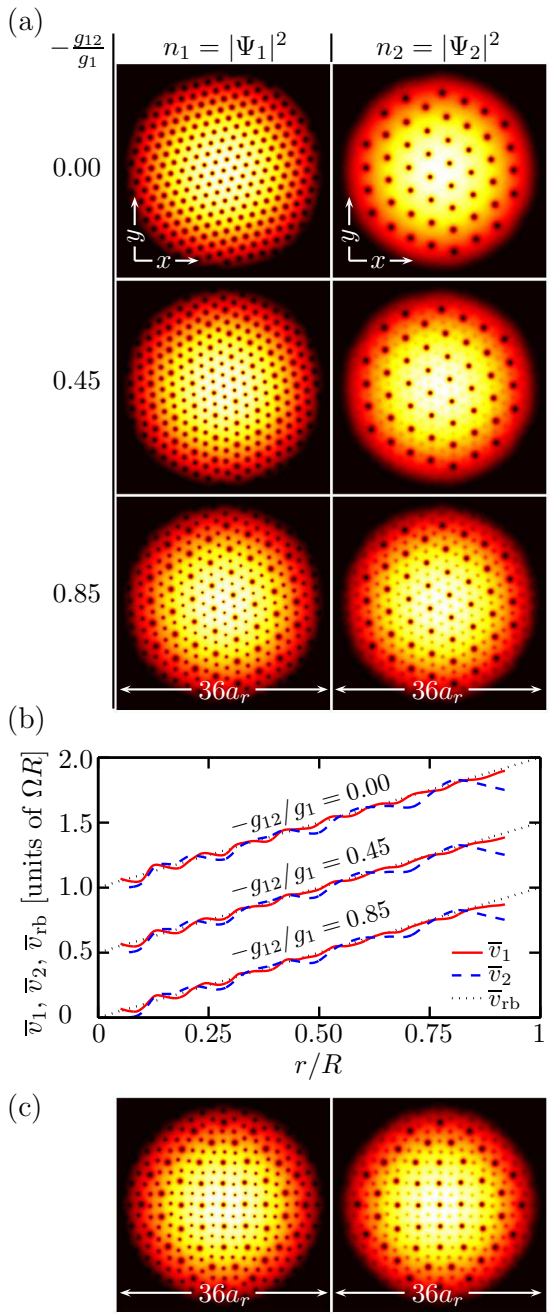}
\caption{\label{fig:rho4profiles} (Color online) Vortex lattices for $m_1/m_2=4$, $\Omega=0.97\,\omega_1$, and $\iaparam = 4000$. (a)~Ground-state densities $n_1$ and $n_2$ at different values of $\sigma$. (b)~Average velocities $\overline{v}_{1}$ and $\overline{v}_{2}$ for the states presented in (a). The curves have been shifted vertically, and $R=17.1\,a_r$. (c) Stationary state at $\sigma=0.85$ having only slightly higher energy than the ground state.
}
\end{figure}

For $\rho=4$, the ground state is found to consist of two triangular lattices as expected [Fig.~\ref{fig:lattices}(d)]. The numerical results are exemplified in Fig.~\ref{fig:rho4profiles}(a), where the ground-state particle densities are presented for $\rho=4$, $\Omega=0.97\,\omega_1$, $\iaparam=4000$, and different values of $\sigma$.  Paired vortices are again found to have larger cores than unpaired ones, and craters develop in $n_2$ at the positions of unpaired vortices in $\Psi_1$ at large values of $\sigma$. Plots of the average velocities $\overline{v}_j$ [Fig.~\ref{fig:rho4profiles}(b)] faithfully reproduce the rigid-body behavior in accordance with the Feynman relation.

Although the triangular lattice is always the ground state for $\rho=4$, its energy at large $\sigma$ turns out to be nearly degenerate with a stationary square-lattice state. An example of such a state is shown in Fig.~\ref{fig:rho4profiles}(c) for $\sigma=0.85$, and it corresponds to the lattice geometry illustrated in Fig.~\ref{fig:lattices}(e). The relative energy difference between this state and the ground state presented in Fig.~\ref{fig:rho4profiles}(a) is $\Delta E / E_0 \approx 4 \times 10^{-5}$. Various metastable structures may thus appear in this parameter region. To study whether the state in Fig.~\ref{fig:rho4profiles}(c) is robust enough to be experimentally realizable, one could solve its elementary excitation spectrum. However, this investigation is left for future work.

\section{Discussion}\label{sc:discussion}

In summary, we have studied vortex lattices in rotating two-component BECs with an atomic mass difference and attractive interaction between the two components. We found that such systems support exotic ground-state vortex configurations, such as square and two-quantum-vortex lattices, when the intercomponent attraction is sufficiently strong and the ratio of atomic masses is suitable. Importantly, the nontriangular geometry or the multiquantum nature of the lattices was not induced by external fields~\cite{Tun2006.PRL97.240402} but emerged as an inherent property of the system. The pairing of vortices between the two components was also observed to affect the vortex core size, paired vortices being larger than unpaired ones.

We also investigated the validity of the Feynman relation in the two-component system and found no significant deviations from it. This should be contrasted with the findings of Barnett \emph{et al.} who studied the attractively interacting two-component BEC for mass ratios only slightly above unity, $1 \leq \rho \leq 1.6$, and concluded that interlocking of the vortices can violate the Feynman relation. However, this anomaly occurred only within a finite distance from the trap center, whereas in the states considered in this work the interlocking behavior extended through the whole system. Therefore, interesting crossover behavior in the regime $1 < \rho < 2$ may occur; this would be relevant to ${}^{87}$Rb--${}^{133}$Cs mixtures~\cite{Pil2009.PRA79.042718,Ler2011.EPJD65.3}, for which $\rho = 1.5$.

Perhaps the most promising system for observing the nontriangular vortex lattices is the two-species BEC of ${}^{41}$K and ${}^{87}$Rb, which has already been realized with tunable intercomponent interactions~\cite{Tha2008.PRL100.210402}. This combination yields the mass ratio $\rho = 2.1$, and we have verified numerically that such a system exhibits ground-state vortex structures similar to those found for $\rho=2$ [Fig.~\ref{fig:rho2profiles}]. In particular, the two-quantum-vortex lattice expected for strong intercomponent attraction would be a rare example of a ground state containing multiply quantized vortices with self-supporting, genuinely empty cores. Since the energy of a vortex increases quadratically with its quantum number, multiply quantized vortices do not usually appear in the ground state.

There are many ways to extend this work in the future. For example, it would be interesting to study the system in the limit of very fast rotation by generalizing the analytical calculations of Mueller and Ho~\cite{Mue2002.PRL88.180403} to dissimilar condensates (cf. also Refs.~\cite{Kec2006.PRA73.023611,Min2009.PRA79.013605}). In the simulations, one could also scan different parameter values more extensively to obtain vortex phase diagrams in the $\left(\sigma,\Omega\right)$ plane~\cite{Kas2003.PRL91.150406,Mas2011.PRA84.033611}. This could be done efficiently by formulating the energetics in terms of the positions of vortices as in Refs.~\cite{Bar2008.NJP10.043030,Aft2012.PRA85.033614}. 

\begin{acknowledgments}
We gratefully acknowledge financial support from the Academy of Finland, the Emil Aaltonen Foundation, the Finnish Cultural Foundation, the KAUTE Foundation, and the V\"ais\"al\"a Foundation. We thank E.~Ruokokoski for insightful discussions and V.~Pietil\"a and S.~M.~M.~Virtanen for their help during the early stages of the work. 
\end{acknowledgments}

\bibliography{tc_manu}

\begin{thebibliography}{10}%
\makeatletter
\providecommand \@ifxundefined [1]{%
 \ifx #1\undefined \expandafter \@firstoftwo
 \else \expandafter \@secondoftwo
\fi
}%
\providecommand \@ifnum [1]{%
 \ifnum #1\expandafter \@firstoftwo
 \else \expandafter \@secondoftwo
\fi
}%
\providecommand \enquote [1]{``#1''}%
\providecommand \bibnamefont  [1]{#1}%
\providecommand \bibfnamefont [1]{#1}%
\providecommand \citenamefont [1]{#1}%
\providecommand\href[0]{\@sanitize\@href}%
\providecommand\@href[1]{\endgroup\@@startlink{#1}\endgroup\@@href}%
\providecommand\@@href[1]{#1\@@endlink}%
\providecommand \@sanitize [0]{\begingroup\catcode`\&12\catcode`\#12\relax}%
\@ifxundefined \pdfoutput {\@firstoftwo}{%
 \@ifnum{\z@=\pdfoutput}{\@firstoftwo}{\@secondoftwo}%
}{%
 \providecommand\@@startlink[1]{\leavevmode\special{html:<a href="#1">}}%
 \providecommand\@@endlink[0]{\special{html:</a>}}%
}{%
 \providecommand\@@startlink[1]{%
  \leavevmode
  \pdfstartlink
   attr{/Border[0 0 1 ]/H/I/C[0 1 1]}%
   user{/Subtype/Link/A<</Type/Action/S/URI/URI(#1)>>}%
  \relax
 }%
 \providecommand\@@endlink[0]{\pdfendlink}%
}%
\providecommand \url  [0]{\begingroup\@sanitize \@url }%
\providecommand \@url [1]{\endgroup\@href {#1}{\urlprefix}}%
\providecommand \urlprefix [0]{URL }%
\providecommand \Eprint[0]{\href }%
\@ifxundefined \urlstyle {%
  \providecommand \doi [1]{doi:\discretionary{}{}{}#1}%
}{%
  \providecommand \doi [0]{doi:\discretionary{}{}{}\begingroup
  \urlstyle{rm}\Url }%
}%
\providecommand \doibase [0]{http://dx.doi.org/}%
\providecommand \Doi[1]{\href{\doibase#1}}%
\providecommand \bibAnnote [3]{%
  \BibitemShut{#1}%
  \begin{quotation}\noindent
    \textsc{Key:}\ #2\\\textsc{Annotation:}\ #3%
  \end{quotation}%
}%
\providecommand \bibAnnoteFile [2]{%
  \IfFileExists{#2}{\bibAnnote {#1} {#2} {\input{#2}}}{}%
}%
\providecommand \typeout [0]{\immediate \write \m@ne }%
\providecommand \selectlanguage [0]{\@gobble}%
\providecommand \bibinfo [0]{\@secondoftwo}%
\providecommand \bibfield [0]{\@secondoftwo}%
\providecommand \translation [1]{[#1]}%
\providecommand \BibitemOpen[0]{}%
\providecommand \bibitemStop [0]{}%
\providecommand \bibitemNoStop [0]{.\EOS\space}%
\providecommand \EOS [0]{\spacefactor3000\relax}%
\providecommand \BibitemShut [1]{\csname bibitem#1\endcsname}%
\bibitem{Ons1949.NCim6Sup2.249}%
  \BibitemOpen
  \bibfield{author}{%
  \bibinfo {author} {\bibfnamefont{L.}~\bibnamefont{Onsager}},\ }%
  \bibfield{journal}{%
  {\bibinfo {journal} {Nuovo Cimento}}\ }%
  \textbf{\bibinfo {volume} {6}}, Suppl. 2,\ \bibinfo {pages} {249} (\bibinfo
  {year} {1949}).%
  \bibAnnoteFile{Stop}{Ons1949.NCim6Sup2.249}%
\bibitem{Fey1955.PLTP1.17}%
  \BibitemOpen
  \bibfield{author}{%
  \bibinfo {author} {\bibfnamefont{R.~P.}\ \bibnamefont{Feynman}},\ }%
  in \emph{\bibinfo {title}{Progress in Low Temperature Physics}},\
  edited by \bibinfo{editor}{\bibfnamefont{C.~J.}\ \bibnamefont{Gorter}} \
  (\bibinfo{publisher}{North-Holland Publishing Company}, Amsterdam, \bibinfo{year}{1955}),\
  Chap.~\bibinfo {chapter} {2}%
  \bibAnnoteFile{Stop}{Fey1955.PLTP1.17}%
\bibitem{Mat1999.PRL83.2498}%
  \BibitemOpen
  \bibfield{author}{%
  \bibinfo {author} {\bibfnamefont{M.~R.}\ \bibnamefont{Matthews}}, \bibinfo
  {author} {\bibfnamefont{B.~P.}\ \bibnamefont{Anderson}}, \bibinfo {author}
  {\bibfnamefont{P.~C.}\ \bibnamefont{Haljan}}, \bibinfo {author}
  {\bibfnamefont{D.~S.}\ \bibnamefont{Hall}}, \bibinfo {author}
  {\bibfnamefont{C.~E.}\ \bibnamefont{Wieman}},\ and\ \bibinfo {author}
  {\bibfnamefont{E.~A.}\ \bibnamefont{Cornell}},\ }%
  \bibfield{journal}{%
  \Doi{10.1103/PhysRevLett.83.2498}{\bibinfo {journal} {Phys. Rev. Lett.}}\ }%
  \textbf{\bibinfo {volume} {83}},\ \bibinfo {pages} {2498} (\bibinfo {year}
  {1999}).%
  \bibAnnoteFile{Stop}{Mat1999.PRL83.2498}%
\bibitem{Mad2000.PRL84.806}%
  \BibitemOpen
  \bibfield{author}{%
  \bibinfo {author} {\bibfnamefont{K.~W.}\ \bibnamefont{Madison}}, \bibinfo
  {author} {\bibfnamefont{F.}~\bibnamefont{Chevy}}, \bibinfo {author}
  {\bibfnamefont{W.}~\bibnamefont{Wohlleben}},\ and\ \bibinfo {author}
  {\bibfnamefont{J.}~\bibnamefont{Dalibard}},\ }%
  \bibfield{journal}{%
  \Doi{10.1103/PhysRevLett.84.806}{\bibinfo {journal} {Phys. Rev. Lett.}}\ }%
  \textbf{\bibinfo {volume} {84}},\ \bibinfo {pages} {806} (\bibinfo {year}
  {2000}).%
  \bibAnnoteFile{Stop}{Mad2000.PRL84.806}%
\bibitem{Abo2001.Sci292.5516}%
  \BibitemOpen
  \bibfield{author}{%
  \bibinfo {author} {\bibfnamefont{J.~R.}\ \bibnamefont{Abo-Shaeer}}, \bibinfo
  {author} {\bibfnamefont{C.}~\bibnamefont{Raman}}, \bibinfo {author}
  {\bibfnamefont{J.~M.}\ \bibnamefont{Vogels}},\ and\ \bibinfo {author}
  {\bibfnamefont{W.}~\bibnamefont{Ketterle}},\ }%
  \bibfield{journal}{%
  \bibinfo {journal} {Science}\ }%
  \textbf{\bibinfo {volume} {292}},\ \bibinfo {pages} {476} (\bibinfo {year}
  {2001}).%
  \bibAnnoteFile{Stop}{Abo2001.Sci292.5516}%
\bibitem{Abr1957.ZETF32.1442}%
  \BibitemOpen
  \bibfield{author}{%
  \bibinfo {author} {\bibfnamefont{A.~A.}\ \bibnamefont{Abrikosov}},\ }%
  \bibfield{journal}{%
  \bibinfo {journal} {Zh. Eksp. Teor. Fiz.}\ }%
  \textbf{\bibinfo {volume} {32}},\ \bibinfo {pages} {1442} (\bibinfo {year}
  {1957}) %
  [\bibinfo {journal} {Sov. Phys. JETP}\
  \textbf{\bibinfo {volume} {5}},\ \bibinfo {pages} {1174} (\bibinfo {year}
  {1957})]%
  \bibAnnoteFile{Stop}{Abr1957.ZETF32.1442}%
\bibitem{Kas2005.IJMPB19.1835}%
  \BibitemOpen
  \bibfield{author}{%
  \bibinfo {author} {\bibfnamefont{K.}~\bibnamefont{Kasamatsu}}, \bibinfo
  {author} {\bibfnamefont{M.}~\bibnamefont{Tsubota}},\ and\ \bibinfo {author}
  {\bibfnamefont{M.}~\bibnamefont{Ueda}},\ }%
  \bibfield{journal}{%
  \bibinfo {journal} {Int. J. Mod. Phys. B}\ }%
  \textbf{\bibinfo {volume} {19}},\ \bibinfo {pages} {1835} (\bibinfo {year}
  {2005}).%
  \bibAnnoteFile{Stop}{Kas2005.IJMPB19.1835}%
\bibitem{Kas2004.PRL93.250406}%
  \BibitemOpen
  \bibfield{author}{%
  \bibinfo {author} {\bibfnamefont{K.}~\bibnamefont{Kasamatsu}}, \bibinfo
  {author} {\bibfnamefont{M.}~\bibnamefont{Tsubota}},\ and\ \bibinfo {author}
  {\bibfnamefont{M.}~\bibnamefont{Ueda}},\ }%
  \bibfield{journal}{%
  \Doi{10.1103/PhysRevLett.93.250406}{\bibinfo {journal} {Phys. Rev. Lett.}}\
  }%
  \textbf{\bibinfo {volume} {93}},\ \bibinfo {pages} {250406} (\bibinfo {year}
  {2004}).%
  \bibAnnoteFile{Stop}{Kas2004.PRL93.250406}%
\bibitem{Kas2005.PRA71.043611}%
  \BibitemOpen
  \bibfield{author}{%
  \bibinfo {author} {\bibfnamefont{K.}~\bibnamefont{Kasamatsu}}, \bibinfo
  {author} {\bibfnamefont{M.}~\bibnamefont{Tsubota}},\ and\ \bibinfo {author}
  {\bibfnamefont{M.}~\bibnamefont{Ueda}},\ }%
  \bibfield{journal}{%
  \Doi{10.1103/PhysRevA.71.043611}{\bibinfo {journal} {Phys. Rev. A}}\ }%
  \textbf{\bibinfo {volume} {71}},\ \bibinfo {pages} {043611} (\bibinfo {year}
  {2005}).%
  \bibAnnoteFile{Stop}{Kas2005.PRA71.043611}%
\bibitem{Kas2009.PRA79.023606}%
  \BibitemOpen
  \bibfield{author}{%
  \bibinfo {author} {\bibfnamefont{K.}~\bibnamefont{Kasamatsu}}\ and\ \bibinfo
  {author} {\bibfnamefont{M.}~\bibnamefont{Tsubota}},\ }%
  \bibfield{journal}{%
  \Doi{10.1103/PhysRevA.79.023606}{\bibinfo {journal} {Phys. Rev. A}}\ }%
  \textbf{\bibinfo {volume} {79}},\ \bibinfo {pages} {023606} (\bibinfo {year}
  {2009}).%
  \bibAnnoteFile{Stop}{Kas2009.PRA79.023606}%
\bibitem{Yan2008.PRA77.033621}%
  \BibitemOpen
  \bibfield{author}{%
  \bibinfo {author} {\bibfnamefont{S.-J.}\ \bibnamefont{Yang}}, \bibinfo
  {author} {\bibfnamefont{Q.-S.}\ \bibnamefont{Wu}}, \bibinfo {author}
  {\bibfnamefont{S.-N.}\ \bibnamefont{Zhang}},\ and\ \bibinfo {author}
  {\bibfnamefont{S.}~\bibnamefont{Feng}},\ }%
  \bibfield{journal}{%
  \Doi{10.1103/PhysRevA.77.033621}{\bibinfo {journal} {Phys. Rev. A}}\ }%
  \textbf{\bibinfo {volume} {77}},\ \bibinfo {pages} {033621} (\bibinfo {year}
  {2008}).%
  \bibAnnoteFile{Stop}{Yan2008.PRA77.033621}%
\bibitem{Mas2011.PRA84.033611}%
  \BibitemOpen
  \bibfield{author}{%
  \bibinfo {author} {\bibfnamefont{P.}~\bibnamefont{Mason}}\ and\ \bibinfo
  {author} {\bibfnamefont{A.}~\bibnamefont{Aftalion}},\ }%
  \bibfield{journal}{%
  \bibinfo {journal} {Phys. Rev. A}\ }%
  \textbf{\bibinfo {volume} {84}},\ \bibinfo {pages} {033611} (\bibinfo {year}
  {2011}).%
  \bibAnnoteFile{Stop}{Mas2011.PRA84.033611}%
\bibitem{Mue2002.PRL88.180403}%
  \BibitemOpen
  \bibfield{author}{%
  \bibinfo {author} {\bibfnamefont{E.~J.}\ \bibnamefont{Mueller}}\ and\
  \bibinfo {author} {\bibfnamefont{T.-L.}\ \bibnamefont{Ho}},\ }%
  \bibfield{journal}{%
  \bibinfo {journal} {Phys. Rev. Lett.}\ }%
  \textbf{\bibinfo {volume} {88}},\ \bibinfo {pages} {180403} (\bibinfo {year}
  {2002}).%
  \bibAnnoteFile{Stop}{Mue2002.PRL88.180403}%
\bibitem{Kas2003.PRL91.150406}%
  \BibitemOpen
  \bibfield{author}{%
  \bibinfo {author} {\bibfnamefont{K.}~\bibnamefont{Kasamatsu}}, \bibinfo
  {author} {\bibfnamefont{M.}~\bibnamefont{Tsubota}},\ and\ \bibinfo {author}
  {\bibfnamefont{M.}~\bibnamefont{Ueda}},\ }%
  \bibfield{journal}{%
  \Doi{10.1103/PhysRevLett.91.150406}{\bibinfo {journal} {Phys. Rev. Lett.}}\
  }%
  \textbf{\bibinfo {volume} {91}},\ \bibinfo {pages} {150406} (\bibinfo {year}
  {2003}).%
  \bibAnnoteFile{Stop}{Kas2003.PRL91.150406}%
\bibitem{Kec2006.PRA73.023611}%
  \BibitemOpen
  \bibfield{author}{%
  \bibinfo {author}
  {\bibfnamefont{M.}~\bibnamefont{Ke\ifmmode~\mbox{\c{c}}\else \c{c}\fi{}eli}}\
  and\ \bibinfo {author} {\bibfnamefont{M.~\"O.}\ \bibnamefont{Oktel}},\ }%
  \bibfield{journal}{%
  \bibinfo {journal} {Phys. Rev. A}\ }%
  \textbf{\bibinfo {volume} {73}},\ \bibinfo {pages} {023611} (\bibinfo {year}
  {2006}).%
  \bibAnnoteFile{Stop}{Kec2006.PRA73.023611}%
\bibitem{Min2009.PRA79.013605}%
  \BibitemOpen
  \bibfield{author}{%
  \bibinfo {author} {\bibfnamefont{M.~P.}\ \bibnamefont{Mink}}, \bibinfo
  {author} {\bibfnamefont{C.~M.}\ \bibnamefont{Smith}},\ and\ \bibinfo {author}
  {\bibfnamefont{R.~A.}\ \bibnamefont{Duine}},\ }%
  \bibfield{journal}{%
  \bibinfo {journal} {Phys. Rev. A}\ }%
  \textbf{\bibinfo {volume} {79}},\ \bibinfo {pages} {013605} (\bibinfo {year}
  {2009}).%
  \bibAnnoteFile{Stop}{Min2009.PRA79.013605}%
\bibitem{Huh2009.PRA80.051601}%
  \BibitemOpen
  \bibfield{author}{%
  \bibinfo {author} {\bibfnamefont{J.~A.~M.}\ \bibnamefont{Huhtam\"aki}},
  \bibinfo {author} {\bibfnamefont{T.~P.}\ \bibnamefont{Simula}}, \bibinfo
  {author} {\bibfnamefont{M.}~\bibnamefont{Kobayashi}},\ and\ \bibinfo {author}
  {\bibfnamefont{K.}~\bibnamefont{Machida}},\ }%
  \bibfield{journal}{%
  \bibinfo {journal} {Phys. Rev. A}\ }%
  \textbf{\bibinfo {volume} {80}},\ \bibinfo {pages} {051601} (\bibinfo {year}
  {2009}).%
  \bibAnnoteFile{Stop}{Huh2009.PRA80.051601}%
\bibitem{Dah2008.PRB78.144510}%
  \BibitemOpen
  \bibfield{author}{%
  \bibinfo {author} {\bibfnamefont{E.~K.}\ \bibnamefont{Dahl}}, \bibinfo
  {author} {\bibfnamefont{E.}~\bibnamefont{Babaev}},\ and\ \bibinfo {author}
  {\bibfnamefont{A.}~\bibnamefont{Sudb\o{}}},\ }%
  \bibfield{journal}{%
  \bibinfo {journal} {Phys. Rev. B}\ }%
  \textbf{\bibinfo {volume} {78}},\ \bibinfo {pages} {144510} (\bibinfo {year}
  {2008}).%
  \bibAnnoteFile{Stop}{Dah2008.PRB78.144510}%
\bibitem{Sch2004.PRL93.210403}%
  \BibitemOpen
  \bibfield{author}{%
  \bibinfo {author} {\bibfnamefont{V.}~\bibnamefont{Schweikhard}}, \bibinfo
  {author} {\bibfnamefont{I.}~\bibnamefont{Coddington}}, \bibinfo {author}
  {\bibfnamefont{P.}~\bibnamefont{Engels}}, \bibinfo {author}
  {\bibfnamefont{S.}~\bibnamefont{Tung}},\ and\ \bibinfo {author}
  {\bibfnamefont{E.~A.}\ \bibnamefont{Cornell}},\ }%
  \bibfield{journal}{%
  \Doi{10.1103/PhysRevLett.93.210403}{\bibinfo {journal} {Phys. Rev. Lett.}}\
  }%
  \textbf{\bibinfo {volume} {93}},\ \bibinfo {pages} {210403} (\bibinfo{year}{2004}).%
  \bibAnnoteFile{Stop}{Sch2004.PRL93.210403}%
\bibitem{Mya1997.PRL78.586}%
  \BibitemOpen
  \bibfield{author}{%
  \bibinfo {author} {\bibfnamefont{C.~J.}\ \bibnamefont{Myatt}}, \bibinfo
  {author} {\bibfnamefont{E.~A.}\ \bibnamefont{Burt}}, \bibinfo {author}
  {\bibfnamefont{R.~W.}\ \bibnamefont{Ghrist}}, \bibinfo {author}
  {\bibfnamefont{E.~A.}\ \bibnamefont{Cornell}},\ and\ \bibinfo {author}
  {\bibfnamefont{C.~E.}\ \bibnamefont{Wieman}},\ }%
  \bibfield{journal}{%
  \Doi{10.1103/PhysRevLett.78.586}{\bibinfo {journal} {Phys. Rev. Lett.}}\ }%
  \textbf{\bibinfo {volume} {78}},\ \bibinfo {pages} {586} (\bibinfo {year}
  {1997}).%
  \bibAnnoteFile{Stop}{Mya1997.PRL78.586}%
\bibitem{Hal1998.PRL81.1539}%
  \BibitemOpen
  \bibfield{author}{%
  \bibinfo {author} {\bibfnamefont{D.~S.}\ \bibnamefont{Hall}}, \bibinfo
  {author} {\bibfnamefont{M.~R.}\ \bibnamefont{Matthews}}, \bibinfo {author}
  {\bibfnamefont{J.~R.}\ \bibnamefont{Ensher}}, \bibinfo {author}
  {\bibfnamefont{C.~E.}\ \bibnamefont{Wieman}},\ and\ \bibinfo {author}
  {\bibfnamefont{E.~A.}\ \bibnamefont{Cornell}},\ }%
  \bibfield{journal}{%
  \Doi{10.1103/PhysRevLett.81.1539}{\bibinfo {journal} {Phys. Rev. Lett.}}\ }%
  \textbf{\bibinfo {volume} {81}},\ \bibinfo {pages} {1539} (\bibinfo {year}
  {1998}).%
  \bibAnnoteFile{Stop}{Hal1998.PRL81.1539}%
\bibitem{Del2001.PRA63.051602}%
  \BibitemOpen
  \bibfield{author}{%
  \bibinfo {author} {\bibfnamefont{G.}~\bibnamefont{Delannoy}}, \bibinfo
  {author} {\bibfnamefont{S.~G.}\ \bibnamefont{Murdoch}}, \bibinfo {author}
  {\bibfnamefont{V.}~\bibnamefont{Boyer}}, \bibinfo {author}
  {\bibfnamefont{V.}~\bibnamefont{Josse}}, \bibinfo {author}
  {\bibfnamefont{P.}~\bibnamefont{Bouyer}},\ and\ \bibinfo {author}
  {\bibfnamefont{A.}~\bibnamefont{Aspect}},\ }%
  \bibfield{journal}{%
  \Doi{10.1103/PhysRevA.63.051602}{\bibinfo {journal} {Phys. Rev. A}}\ }%
  \textbf{\bibinfo {volume} {63}},\ \bibinfo {pages} {051602} (\bibinfo {year}
  {2001}).%
  \bibAnnoteFile{Stop}{Del2001.PRA63.051602}%
\bibitem{And2009.PRA80.023603}%
  \BibitemOpen
  \bibfield{author}{%
  \bibinfo {author} {\bibfnamefont{R.~P.}\ \bibnamefont{Anderson}}, \bibinfo
  {author} {\bibfnamefont{C.}~\bibnamefont{Ticknor}}, \bibinfo {author}
  {\bibfnamefont{A.~I.}\ \bibnamefont{Sidorov}},\ and\ \bibinfo {author}
  {\bibfnamefont{B.~V.}\ \bibnamefont{Hall}},\ }%
  \bibfield{journal}{%
  \Doi{10.1103/PhysRevA.80.023603}{\bibinfo {journal} {Phys. Rev. A}}\ }%
  \textbf{\bibinfo {volume} {80}},\ \bibinfo {pages} {023603} (\bibinfo {year}
  {2009}).%
  \bibAnnoteFile{Stop}{And2009.PRA80.023603}%
\bibitem{Bap2008.PRL101.040402}%
  \BibitemOpen
  \bibfield{author}{%
  \bibinfo {author} {\bibfnamefont{S.~B.}\ \bibnamefont{Papp}}, \bibinfo
  {author} {\bibfnamefont{J.~M.}\ \bibnamefont{Pino}},\ and\ \bibinfo {author}
  {\bibfnamefont{C.~E.}\ \bibnamefont{Wieman}},\ }%
  \bibfield{journal}{%
  \Doi{10.1103/PhysRevLett.101.040402}{\bibinfo {journal} {Phys. Rev. Lett.}}\
  }%
  \textbf{\bibinfo {volume} {101}},\ \bibinfo {pages} {040402} (\bibinfo {year}
  {2008}).%
  \bibAnnoteFile{Stop}{Bap2008.PRL101.040402}%
\bibitem{Sug2011.PRA84.011610}%
  \BibitemOpen
  \bibfield{author}{%
  \bibinfo {author} {\bibfnamefont{S.}~\bibnamefont{Sugawa}}, \bibinfo {author}
  {\bibfnamefont{R.}~\bibnamefont{Yamazaki}}, \bibinfo {author}
  {\bibfnamefont{S.}~\bibnamefont{Taie}},\ and\ \bibinfo {author}
  {\bibfnamefont{Y.}~\bibnamefont{Takahashi}},\ }%
  \bibfield{journal}{%
  \Doi{10.1103/PhysRevA.84.011610}{\bibinfo {journal} {Phys. Rev. A}}\ }%
  \textbf{\bibinfo {volume} {84}},\ \bibinfo {pages} {011610} (\bibinfo {year}
  {2011}).%
  \bibAnnoteFile{Stop}{Sug2011.PRA84.011610}%
\bibitem{Fer2002.PRL89.053202}%
  \BibitemOpen
  \bibfield{author}{%
  \bibinfo {author} {\bibfnamefont{G.}~\bibnamefont{Ferrari}}, \bibinfo
  {author} {\bibfnamefont{M.}~\bibnamefont{Inguscio}}, \bibinfo {author}
  {\bibfnamefont{W.}~\bibnamefont{Jastrzebski}}, \bibinfo {author}
  {\bibfnamefont{G.}~\bibnamefont{Modugno}}, \bibinfo {author}
  {\bibfnamefont{G.}~\bibnamefont{Roati}},\ and\ \bibinfo {author}
  {\bibfnamefont{A.}~\bibnamefont{Simoni}},\ }%
  \bibfield{journal}{%
  \Doi{10.1103/PhysRevLett.89.053202}{\bibinfo {journal} {Phys. Rev. Lett.}}\
  }%
  \textbf{\bibinfo {volume} {89}},\ \bibinfo {pages} {053202} (\bibinfo {year}
  {2002}).%
  \bibAnnoteFile{Stop}{Fer2002.PRL89.053202}%
\bibitem{Mod2002.PRL89.190404}%
  \BibitemOpen
  \bibfield{author}{%
  \bibinfo {author} {\bibfnamefont{G.}~\bibnamefont{Modugno}}, \bibinfo
  {author} {\bibfnamefont{M.}~\bibnamefont{Modugno}}, \bibinfo {author}
  {\bibfnamefont{F.}~\bibnamefont{Riboli}}, \bibinfo {author}
  {\bibfnamefont{G.}~\bibnamefont{Roati}},\ and\ \bibinfo {author}
  {\bibfnamefont{M.}~\bibnamefont{Inguscio}},\ }%
  \bibfield{journal}{%
  \bibinfo {journal} {Phys. Rev. Lett.}\ }%
  \textbf{\bibinfo {volume} {89}},\ \bibinfo {pages} {190404} (\bibinfo {year}
  {2002}).%
  \bibAnnoteFile{Stop}{Mod2002.PRL89.190404}%
\bibitem{Cat2008.PRA77.011603}%
  \BibitemOpen
  \bibfield{author}{%
  \bibinfo {author} {\bibfnamefont{J.}~\bibnamefont{Catani}}, \bibinfo {author}
  {\bibfnamefont{L.}~\bibnamefont{De~Sarlo}}, \bibinfo {author}
  {\bibfnamefont{G.}~\bibnamefont{Barontini}}, \bibinfo {author}
  {\bibfnamefont{F.}~\bibnamefont{Minardi}},\ and\ \bibinfo {author}
  {\bibfnamefont{M.}~\bibnamefont{Inguscio}},\ }%
  \bibfield{journal}{%
  \bibinfo {journal} {Phys. Rev. A}\ }%
  \textbf{\bibinfo {volume} {77}},\ \bibinfo {pages} {011603} (\bibinfo {year}
  {2008}).%
  \bibAnnoteFile{Stop}{Cat2008.PRA77.011603}%
\bibitem{Tha2008.PRL100.210402}%
  \BibitemOpen
  \bibfield{author}{%
  \bibinfo {author} {\bibfnamefont{G.}~\bibnamefont{Thalhammer}}, \bibinfo
  {author} {\bibfnamefont{G.}~\bibnamefont{Barontini}}, \bibinfo {author}
  {\bibfnamefont{L.}~\bibnamefont{De~Sarlo}}, \bibinfo {author}
  {\bibfnamefont{J.}~\bibnamefont{Catani}}, \bibinfo {author}
  {\bibfnamefont{F.}~\bibnamefont{Minardi}},\ and\ \bibinfo {author}
  {\bibfnamefont{M.}~\bibnamefont{Inguscio}},\ }%
  \bibfield{journal}{%
  \Doi{10.1103/PhysRevLett.100.210402}{\bibinfo {journal} {Phys. Rev. Lett.}}\
  }%
  \textbf{\bibinfo {volume} {100}},\ \bibinfo {pages} {210402} (\bibinfo {year}
  {2008}).%
  \bibAnnoteFile{Stop}{Tha2008.PRL100.210402}%
\bibitem{Aik2009.NJP11.055035}%
  \BibitemOpen
  \bibfield{author}{%
  \bibinfo {author} {\bibfnamefont{K.}~\bibnamefont{Aikawa}}, \bibinfo {author}
  {\bibfnamefont{D.}~\bibnamefont{Akamatsu}}, \bibinfo {author}
  {\bibfnamefont{J.}~\bibnamefont{Kobayashi}}, \bibinfo {author}
  {\bibfnamefont{M.}~\bibnamefont{Ueda}}, \bibinfo {author}
  {\bibfnamefont{T.}~\bibnamefont{Kishimoto}},\ and\ \bibinfo {author}
  {\bibfnamefont{S.}~\bibnamefont{Inouye}},\ }%
  \bibfield{journal}{%
  \bibinfo {journal} {New J. Phys.}\ }%
  \textbf{\bibinfo {volume} {11}},\ \bibinfo {pages} {055035} (\bibinfo {year}
  {2009}).%
  \bibAnnoteFile{Stop}{Aik2009.NJP11.055035}%
\bibitem{Ler2011.EPJD65.3}%
  \BibitemOpen
  \bibfield{author}{%
  \bibinfo {author} {\bibfnamefont{A.}~\bibnamefont{Lercher}}, \bibinfo
  {author} {\bibfnamefont{T.}~\bibnamefont{Takekoshi}}, \bibinfo {author}
  {\bibfnamefont{M.}~\bibnamefont{Debatin}}, \bibinfo {author}
  {\bibfnamefont{B.}~\bibnamefont{Schuster}}, \bibinfo {author}
  {\bibfnamefont{R.}~\bibnamefont{Rameshan}}, \bibinfo {author}
  {\bibfnamefont{F.}~\bibnamefont{Ferlaino}}, \bibinfo {author}
  {\bibfnamefont{R.}~\bibnamefont{Grimm}},\ and\ \bibinfo {author}
  {\bibfnamefont{H.-C.}~\bibnamefont{N\"agerl}},\ }%
  \bibfield{journal}{%
  \bibinfo {journal} {Eur. Phys. J. D}\ }%
  \textbf{\bibinfo {volume} {65}},\ \bibinfo {pages} {3} (\bibinfo {year}
  {2011}).%
  \bibAnnoteFile{Stop}{Ler2011.EPJD65.3}%
\bibitem{Bar2008.NJP10.043030}%
  \BibitemOpen
  \bibfield{author}{%
  \bibinfo {author} {\bibfnamefont{R.}~\bibnamefont{Barnett}}, \bibinfo
  {author} {\bibfnamefont{G.}~\bibnamefont{Refael}}, \bibinfo {author}
  {\bibfnamefont{M.~A.}\ \bibnamefont{Porter}},\ and\ \bibinfo {author}
  {\bibfnamefont{H.~P.}\ \bibnamefont{B\"uchler}},\ }%
  \bibfield{journal}{%
  \bibinfo {journal} {New J. Phys.}\ }%
  \textbf{\bibinfo {volume} {10}},\ \bibinfo {pages} {043030} (\bibinfo {year}
  {2008}).%
  \bibAnnoteFile{Stop}{Bar2008.NJP10.043030}%
\bibitem{Bar2010.NJP12.043004}%
  \BibitemOpen
  \bibfield{author}{%
  \bibinfo {author} {\bibfnamefont{R.}~\bibnamefont{Barnett}}, \bibinfo
  {author} {\bibfnamefont{E.}~\bibnamefont{Chen}},\ and\ \bibinfo {author}
  {\bibfnamefont{G.}~\bibnamefont{Refael}},\ }%
  \bibfield{journal}{%
  \bibinfo {journal} {New J. Phys.}\ }%
  \textbf{\bibinfo {volume} {12}},\ \bibinfo {pages} {043004} (\bibinfo {year}
  {2010}).%
  \bibAnnoteFile{Stop}{Bar2010.NJP12.043004}%
  \bibitem{Note_trap}%
  \BibitemOpen
  \bibinfo {note} {In the case of strong harmonic confinement in the $z$ direction with trapping frequencies $\left\{\omega_{z,j}\right\}$, the coupling constants are given by $g_j = \sqrt{8 \pi}\hbar^2 a_j / m_j a_{z,j}$ and $g_{12} = 2 \sqrt{\pi} \hbar^2 a_{12} / m_{12} \left( a_{z,1}^2 + a_{z,2}^2\right)^{1/2}$, where $a_{z,j}=\sqrt{\hbar/m_j \omega_{z,j}}$ and $m_{12}=m_1 m_2 /\left( m_1+m_2 \right)$ is the reduced mass.}%
  \bibAnnoteFile{Stop}{Note_trap}%
\bibitem{Fet2009.RMP81.647}%
  \BibitemOpen
  \bibfield{author}{%
  \bibinfo {author} {\bibfnamefont{A.~L.}\ \bibnamefont{Fetter}},\ }%
  \bibfield{journal}{%
  \bibinfo {journal} {Rev. Mod. Phys.}\ }%
  \textbf{\bibinfo {volume} {81}},\ \bibinfo {pages} {647} (\bibinfo {year}
  {2009}).%
  \bibAnnoteFile{Stop}{Fet2009.RMP81.647}%
\bibitem{Cor2009.PRA80.033609}%
  \BibitemOpen
  \bibfield{author}{%
  \bibinfo {author} {\bibfnamefont{I.}~\bibnamefont{Corro}}, \bibinfo {author}
  {\bibfnamefont{R.~G.}\ \bibnamefont{Scott}},\ and\ \bibinfo {author}
  {\bibfnamefont{A.~M.}\ \bibnamefont{Martin}},\ }%
  \bibfield{journal}{%
  \bibinfo {journal} {Phys. Rev. A}\ }%
  \textbf{\bibinfo {volume} {80}},\ \bibinfo {pages} {033609} (\bibinfo {year}
  {2009}).%
  \bibAnnoteFile{Stop}{Cor2009.PRA80.033609}%
  \bibitem{Note_mqv}%
  \BibitemOpen
  \bibinfo {note} {Vortices with $k > 1$ quanta are not expected in the ground state, because their energy scales as $k^2$. Nevertheless, two-quantum vortices are observed for $\rho=2$ when the intercomponent attraction is very strong (Sec.~\ref{sc:results}).}%
  \bibAnnoteFile{Stop}{Note_mqv}%
\bibitem{Chi2010.RMP82.1225}%
  \BibitemOpen
  \bibfield{author}{%
  \bibinfo {author} {\bibfnamefont{C.}~\bibnamefont{Chin}}, \bibinfo {author}
  {\bibfnamefont{R.}~\bibnamefont{Grimm}}, \bibinfo {author}
  {\bibfnamefont{P.}~\bibnamefont{Julienne}},\ and\ \bibinfo {author}
  {\bibfnamefont{E.}~\bibnamefont{Tiesinga}},\ }%
  \bibfield{journal}{%
  \Doi{10.1103/RevModPhys.82.1225}{\bibinfo {journal} {Rev. Mod. Phys.}}\ }%
  \textbf{\bibinfo {volume} {82}},\ \bibinfo {pages} {1225} (\bibinfo {year}
  {2010}).%
  \bibAnnoteFile{Stop}{Chi2010.RMP82.1225}%
\bibitem{Sta2004.PRL93.143001}%
  \BibitemOpen
  \bibfield{author}{%
  \bibinfo {author} {\bibfnamefont{C.~A.}\ \bibnamefont{Stan}}, \bibinfo
  {author} {\bibfnamefont{M.~W.}\ \bibnamefont{Zwierlein}}, \bibinfo {author}
  {\bibfnamefont{C.~H.}\ \bibnamefont{Schunck}}, \bibinfo {author}
  {\bibfnamefont{S.~M.~F.}\ \bibnamefont{Raupach}},\ and\ \bibinfo {author}
  {\bibfnamefont{W.}~\bibnamefont{Ketterle}},\ }%
  \bibfield{journal}{%
  \Doi{10.1103/PhysRevLett.93.143001}{\bibinfo {journal} {Phys. Rev. Lett.}}\
  }%
  \textbf{\bibinfo {volume} {93}},\ \bibinfo {pages} {143001} (\bibinfo {year}
  {2004}).%
  \bibAnnoteFile{Stop}{Sta2004.PRL93.143001}%
\bibitem{Ino2004.PRL93.183201}%
  \BibitemOpen
  \bibfield{author}{%
  \bibinfo {author} {\bibfnamefont{S.}~\bibnamefont{Inouye}}, \bibinfo {author}
  {\bibfnamefont{J.}~\bibnamefont{Goldwin}}, \bibinfo {author}
  {\bibfnamefont{M.~L.}\ \bibnamefont{Olsen}}, \bibinfo {author}
  {\bibfnamefont{C.}~\bibnamefont{Ticknor}}, \bibinfo {author}
  {\bibfnamefont{J.~L.}\ \bibnamefont{Bohn}},\ and\ \bibinfo {author}
  {\bibfnamefont{D.~S.}\ \bibnamefont{Jin}},\ }%
  \bibfield{journal}{%
  \Doi{10.1103/PhysRevLett.93.183201}{\bibinfo {journal} {Phys. Rev. Lett.}}\
  }%
  \textbf{\bibinfo {volume} {93}},\ \bibinfo {pages} {183201} (\bibinfo {year}
  {2004}).%
  \bibAnnoteFile{Stop}{Ino2004.PRL93.183201}%
\bibitem{Pil2009.PRA79.042718}%
  \BibitemOpen
  \bibfield{author}{%
  \bibinfo {author} {\bibfnamefont{K.}~\bibnamefont{Pilch}}, \bibinfo {author}
  {\bibfnamefont{A.~D.}\ \bibnamefont{Lange}}, \bibinfo {author}
  {\bibfnamefont{A.}~\bibnamefont{Prantner}}, \bibinfo {author}
  {\bibfnamefont{G.}~\bibnamefont{Kerner}}, \bibinfo {author}
  {\bibfnamefont{F.}~\bibnamefont{Ferlaino}}, \bibinfo {author}
  {\bibfnamefont{H.-C.}\ \bibnamefont{N\"agerl}},\ and\ \bibinfo {author}
  {\bibfnamefont{R.}~\bibnamefont{Grimm}},\ }%
  \bibfield{journal}{%
  \bibinfo {journal} {Phys. Rev. A}\ }%
  \textbf{\bibinfo {volume} {79}},\ \bibinfo {pages} {042718} (\bibinfo {year}
  {2009}).%
  \bibAnnoteFile{Stop}{Pil2009.PRA79.042718}%
\bibitem{Dah2008.PRL101.255301}%
  \BibitemOpen
  \bibfield{author}{%
  \bibinfo {author} {\bibfnamefont{E.~K.}\ \bibnamefont{Dahl}}, \bibinfo
  {author} {\bibfnamefont{E.}~\bibnamefont{Babaev}},\ and\ \bibinfo {author}
  {\bibfnamefont{A.}~\bibnamefont{Sudb\o{}}},\ }%
  \bibfield{journal}{%
  \bibinfo {journal} {Phys. Rev. Lett.}\ }%
  \textbf{\bibinfo {volume} {101}},\ \bibinfo {pages} {255301} (\bibinfo {year}
  {2008}).%
  \bibAnnoteFile{Stop}{Dah2008.PRL101.255301}%
  \bibitem{Tun2006.PRL97.240402}%
  \BibitemOpen
  \bibfield{author}{%
  \bibinfo {author} {\bibfnamefont{S.}\ \bibnamefont{Tung}}, \bibinfo
  {author} {\bibfnamefont{V.}\ \bibnamefont{Schweikhard}},\ and\ \bibinfo {author}
  {\bibfnamefont{E.~A.}\ \bibnamefont{Cornell}},\ }%
  \bibfield{journal}{%
  \Doi{10.1103/PhysRevLett.97.240402}{\bibinfo {journal} {Phys. Rev. Lett.}}\
  }%
  \textbf{\bibinfo {volume} {97}},\ \bibinfo {pages} {240402} (\bibinfo {year}
  {2006}).%
  \bibAnnoteFile{Stop}{Tun2006.PRL97.240402}%
\bibitem{Aft2012.PRA85.033614}%
  \BibitemOpen
  \bibfield{author}{%
  \bibinfo {author} {\bibfnamefont{A.}\ \bibnamefont{Aftalion}},  
  \bibinfo {author} {\bibfnamefont{P.}~\bibnamefont{Mason}},\ and\ \bibinfo
  {author} {\bibfnamefont{J.}~\bibnamefont{Wei}},\ }%
  \bibfield{journal}{%
  \bibinfo {journal} {Phys. Rev. A}\ }%
  \textbf{\bibinfo {volume} {85}},\ \bibinfo {pages} {033614} (\bibinfo {year}
  {2012}).%
  \bibAnnoteFile{Stop}{Aft2012.PRA85.033614}%
\end{thebibliography}%
\end{document}